%% file: IEEE-conference-template-062824.tex
\documentclass[conference]{IEEEtran}
\IEEEoverridecommandlockouts

\usepackage{amsmath,graphicx}
\usepackage{multicol}
\usepackage{multirow}
\usepackage{threeparttable,booktabs}
\usepackage{booktabs}
\usepackage{cite}

\usepackage{bbm}
\usepackage{amsmath}
\usepackage{dsfont}
\usepackage{xcolor}
\usepackage{algorithm}
\usepackage{arydshln}
\usepackage{resizegather}
\usepackage{url}
\usepackage{makecell}
\usepackage{lineno,hyperref}

\usepackage{threeparttable,booktabs}

\usepackage{graphicx}               
\usepackage{amsmath,amssymb}
\usepackage[noend]{algpseudocode}
\usepackage{amssymb}
\usepackage{pifont}

\usepackage[utf8]{inputenc}
\usepackage[english]{babel}
\usepackage{siunitx}
\usepackage{multirow}
\usepackage{caption}

\usepackage{tabstackengine}
\TABstackMath
\setstackgap{S}{2pt}

\def\ie{\emph{i.e., }}

\def\eg{\emph{e.g., }}

\def\BibTeX{{\rm B\kern-.05em{\sc i\kern-.025em b}\kern-.08em
    T\kern-.1667em\lower.7ex\hbox{E}\kern-.125emX}}

\begin{document}

\title{Full-Rank No More: Low-Rank Weight Training for Modern Speech Recognition Models}

\author{
Adriana Fernandez-Lopez$^{1}$, Shiwei Liu$^{2}$, Lu Yin$^{3}$, Stavros Petridis$^{1,4}$, Maja Pantic$^{1,4}$ \\[0.5ex]$^1$Meta, UK \quad $^2$University of Oxford, UK\quad $^3$University of Surrey, UK\quad
$^4$Imperial College London, UK \thanks{All experimentation and processing were conducted by the authors at Imperial College London and University of Surrey on Imperial College London and University of Surrey’s servers.}
}

\maketitle

\begin{abstract}
This paper investigates the under-explored area of low-rank weight training for large-scale Conformer-based speech recognition models from scratch. Our study demonstrates the viability of this training paradigm for such models, yielding several notable findings. Firstly, we discover that applying a low-rank structure exclusively to the attention modules can unexpectedly enhance performance, even with a significant rank reduction of 12\%. In contrast, feed-forward layers present greater challenges, as they begin to exhibit performance degradation with a moderate 50\% rank reduction. Furthermore, we find that both initialization and layer-wise rank assignment play critical roles in successful low-rank training. Specifically, employing SVD initialization and linear layer-wise rank mapping significantly boosts the efficacy of low-rank weight training. Building on these insights, we introduce the Low-Rank Speech Model from Scratch (LR-SMS), an approach that achieves performance parity with full-rank training while delivering substantial reductions in parameters count (by at least 2$\times$), and training time speedups (by 1.3$\times$ for ASR and 1.15$\times$ for AVSR).

\end{abstract}

\begin{IEEEkeywords}
low-rank training, speech recognition 
\end{IEEEkeywords}

\input{sections/introduction}

\input{sections/method}

\input{sections/experimental_setup}

\input{sections/results}

\input{sections/conclusions}

\clearpage
\bibliographystyle{IEEEbib}
\bibliography{refs}

\end{document}

%% file: sections/introduction.tex
\section{Introduction}
\label{sec:intro}

Training large machine learning (ML) models for computer vision, natural language processing, and speech recognition has become increasingly challenging due to the exponential growth in the number of parameters, scaling from millions \cite{devlin2018bert} to billions \cite{achiam2023gpt, touvron2023llama}, and even reaching trillions \cite{fedus2022switch}. To expedite the training process and alleviate memory constraints, researchers have explored numerous techniques aimed at reducing the parameter count of modern neural networks.

One avenue of research has focused on the design of resource-efficient networks, exemplified by models such as MobileNet \cite{koumparoulis2019mobilipnet} and EfficientNet \cite{koumparoulis2022accurate}. While these models reduce size, they often come at the expense of performance. Another widely adopted approach is network pruning, which seeks to eliminate redundant or less significant components within a model. Unstructured pruning, for example, removes individual weights and achieves high compression rates, but is not hardware-friendly \cite{han2015deep, liu2021we,fernandez2024msrs}. Conversely, structured pruning, which removes entire structures such as channels or attention heads, offers real speedups on common hardware, though it typically achieves lower compression ratios \cite{yin2024dynamic}.

\input{figures/encoder_block}

Recently, low-rank matrix factorization has garnered significant attention, particularly with the advent of Low-Rank Adaptors (LoRA) \cite{hu2021lora}. This approach dramatically reduces memory usage and computation time, making it highly effective for parameter-efficient fine-tuning of large language models (LLMs). Following this development, the ML community has increasingly focused on low-rank fine-tuning, leading to the emergence of several enhanced methods \cite{zhang2023adalora, qiu2023controlling, liu2024dora,li2024owlore,dong2024adarank}. Given that base models are pre-trained on massive corpora, it is unsurprising that a few low-rank adaptors can yield satisfactory fine-tuning performance.

However, when it comes to training large-scale neural networks from scratch, directly applying low-rank structures often compromises performance \cite{wei2024investigating}. Some degree of full-rank training appears necessary to achieve comparable performance, as evidenced by approaches that employ short full-rank ``warm-up'' phases before transitioning to low-rank training \cite{lialin2023relora, wang2023cuttlefish, wei2024investigating, jaiswal2024WeLore}, or strategies that maintain full-rank weight matrices while applying low-rank gradients \cite{zhao2024galore, zhang2024q, han2024sltrain}. Despite these advancements, it remains unclear whether modern networks can be trained directly with low-rank weights from scratch, without compromising performance, and while retaining both parameter and memory efficiency. Additionally, the potential of low-rank architectures has been underexplored in the context of speech recognition, where prior efforts have been confined to relatively small model sizes, typically under 100 million parameters \cite{winata2020lightweight,povey2018semi, chien2017deep,hernandez2023sharing,mcgraw2016personalized}. The efficacy of low-rank training in large-scale speech recognition models remains largely unexamined, especially with Conformer-based models \cite{ma2023auto, gulati2020conformer}.

To address this gap, we redirect our attention from pursuing state-of-the-art performance to systematically investigate low-rank training from scratch for large-scale Automatic Speech Recognition (ASR) and Audio-Visual Speech Recognition (AVSR) models. Our study begins with an in-depth analysis of low-rank emergence during full-rank weight training, revealing that the entire model gradually learns a low-rank structure during training, and some layers exhibit lower ranks than others, as shown in Figure \ref{fig:encoderASR}. Encouraged by these findings, we explore the feasibility of training models directly using low-rank approximations from the outset. Our results reveal several key insights:

\textbf{(i)} Surprisingly, the performance of ASR models can be improved by applying low-rank constraints exclusively to the multi-headed self-attention (MHSA) modules, even with a significant reduction in rank (e.g., 12\%). Conversely, the feed-forward network (FFN) layers are harder to train with low rank than MHSA layers. Low-rank FFN configurations suffer a notable performance drop with a 50\% rank reduction. 
    
\textbf{(ii)} We identified two techniques that enhance low-rank training: (1) SVD Initialization: initializing low-rank matrices with SVD decomposition; (2) Linear layer-wise rank: assigning a linearly increasing rank across the entire model, from top to bottom, including both MHSA and FFN layers. 
    
\textbf{(iii)} Building on these findings, we propose the Low-Rank Speech Model from Scratch (LR-SMS), a simple but effective approach that enables the training of large-scale speech models with low-rank weights without compromising performance. LR-SMS matches the performance of full-rank training while achieving substantial reductions in parameters count (by 2$\times$), and training time (by 1.3$\times$ for ASR and 1.15$\times$ for AVSR).

%% file: figures/encoder_block.tex
\begin{figure}[tb]
  \centering
  \includegraphics[width=0.95\columnwidth, trim=0cm 1cm 0cm 0.2cm, clip]{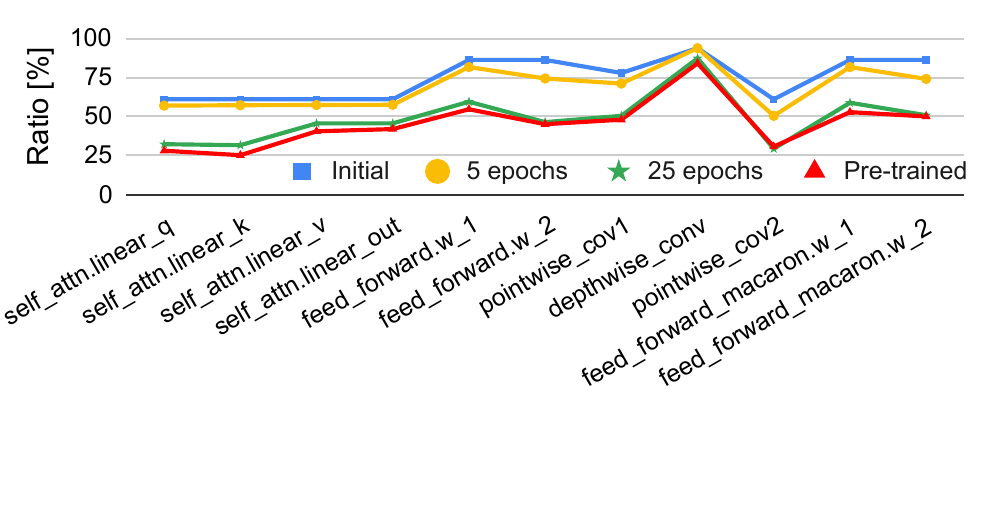}
  \vspace{-0.8cm}
  \captionsetup{font=scriptsize}
  \caption{Close-up view of the topology of a block of the ASR Conformer Encoder at different training stages. Ratio: number of singular vectors required to approximate 95\% of each weight matrix compared to the total number of singular values. A large ratio indicates that most singular vectors are necessary, while a small means few are needed.}
  \label{fig:encoderASR}
  \vspace{-0.6cm}
\end{figure}

%% file: sections/method.tex
\section{Methodology}
\label{sec:method}
\input{figures/architecture}

Let us examine a widely used architecture for ASR/AVSR, as described in \cite{ma2023auto}. This architecture accepts audio or audio-visual data as input and generates the corresponding grapheme transcription. We'll focus on ASR for illustration purposes, but the concepts can be applied to any speech model. Figure \ref{fig:architecture}-(a) illustrates an ASR model consisting of a ResNet frontend, a Conformer encoder and a Transformer decoder that is jointly trained using Connectionist Temporal Classification (CTC loss) and Cross-Entropy (CE Loss). Our objective is to minimize the model's size and accelerate the training process while preserving its complexity and performance. We choose to low-rank the linear layers in the Conformer encoder and Transformer decoder, which includes the linear projections in both FFNs and MHSA. Illustrations of these parameterized modules are provided in Figures \ref{fig:architecture}-(b) and \ref{fig:architecture}-(d) for the Conformer encoder, but the same applies to the Transformer decoder.

\vspace{-0.1cm}
\subsection{Simplifying linear layers with low-rank factorization}
\label{sec:linear_factorization}

A linear layer processes an $n$-dimensional input vector $x$ and produces an $m$-dimensional vector $z=Wx$. Our aim is to factorize the weight matrix $W \in \mathbb{R}^{m \times n}$ into a lower-rank representation, specifically, a product of two matrices $U \in \mathbb{R}^{m \times r}$ and $V^T \in \mathbb{R}^{r \times n}$, where $r$ is much smaller than both $m$ and $n$. Therefore, the parameterized linear layer is given as $z \approx U(V^Tx)$ (shown in Fig. \ref{fig:architecture}-(c)). Computation and memory costs are reduced from $O(mn$) to $O(r(m+n))$. 

\vspace{-0.1em}
\subsection{Effective initialization}
\label{sec:svd_init}

The initialization of model parameters is a crucial step in achieving optimal performance, and it requires careful consideration. Glorot et al. \cite{glorot2010understanding} and He et al. \cite{he2015delving} proposed two widely used initialization methods for FFNs that adhere to two key principles: i) the mean of the activation should be zero; and ii) the variance of the activation should remain consistent across every layer. By ensuring a zero mean and maintaining the variance of the input to each layer, these methods guarantee a stable signal that neither explodes nor vanishes. To achieve low-rank training, it is essential to maintain the variance of the original matrix consistently across each layer after decomposing it into the product of two matrices, \ie  var(W) $\approx$ var(U$V^T$). Several attempts have been made to preserve these principles, as discussed in \cite{saada2023initialisation, khodak2021initialization, ioannou2015training}. We follow \cite{khodak2021initialization}, who presented \textit{spectral initialization} as a technique that simulates the behavior of existing initialization. To initialize the model weights, we follow conventional initializations as described in \cite{glorot2010understanding} and \cite{he2015delving}, and then for each factorized layer, we use the spectral SVD initialization scheme  \cite{khodak2021initialization}, as shown in Eq. (\ref{eq:svd_init}).
\vspace{-0.3cm}

\begin{small}
\begin{equation}
    {\text{SVD}}_r(W) = \hat{U}_{:r}\Sigma_r\hat{V}_{:r}^T, \;  U = \hat{U}_{:r}\sqrt{\Sigma_r}, \; V^T = \sqrt{\Sigma_r}\hat{V}^T_{:r} 
    \label{eq:svd_init}
\end{equation}
\end{small}

\subsection{Optimizing rank for different layers}

Determining distinct ranks for different layers is optimal, as layers contribute unequally to the overall model performance \cite{fernandez2024msrs}. However, identifying the appropriate rank for each layer remains a significant challenge. 

Formally, assuming that our model $\mathcal{M}$ contains $L$ layers, we intend to factorize a subset $\mathcal{S}$ of them. For each layer $l \in \mathcal{S}$, we assign a scaling factor $\alpha_l \in [0, 1] $. The rank $r_l$ for each layer is then determined by multiplying the minimum dimension of the weight matrix $W_l \in \mathbb{R}^{m \times n}$ by the scale factor $\alpha_l$, \ie $ r_l = \alpha_l \cdot \min\{m, n\}$,  for all $l \in \mathcal{S}$. Smaller $\alpha$ represents a lower rank.

\subsubsection{Uniform Layer-wise Rank} Previous works usually assign a uniform rank for all layers without considering layer importance, \ie $\alpha_l = \alpha$, for all $l \in \mathcal{S}$, \eg \cite{winata2020lightweight,hernandez2023sharing}. However, given the fact that layers in neural networks are not equally important, this rank assignment may lead to suboptimal performance.

\label{sec:uniform_rank}

\input{figures/linear_mapping}

\subsubsection{Linear Layer-wise Rank} 
\label{sec:linear_rank}

In fact, different layers within a model can exhibit distinct low-rank patterns. To illustrate this, we analyze the emergence of low-rank structures during the training of a Conformer block. As depicted in Fig. \ref{fig:encoderASR}, the entire model gradually learns a low-rank structure as training progresses, with some layers consistently displaying lower ranks than others. This observation naturally prompts further investigation into the behavior of these layers across varying depths. Figures \ref{fig:linear_mapping}-(a) and \ref{fig:linear_mapping}-(b) present the compression ratios of Conformer and Transformer blocks in the ASR model pre-trained on the Librispeech dataset across different depths. Notably, a near-linear trend is observed, indicating that early blocks generally have lower ranks compared to late blocks.

Inspired by this observation, we propose a \textit{Linear Layer-Wise Rank}, which is a simple but effective rank assignment approach. For each layer $l \in$ $\mathcal{S}$, we assign a linear scaling factor $\alpha_l$ that increases its value through depth, where:
\begin{equation}
\label{eq:linear_rank}
    \alpha_l = b \cdot ( \alpha^f - \alpha^i ) / B  + \alpha^i
\end{equation}
[$\alpha^i, \alpha^f]$ represents the range of scaling factors for the submodel, with $\alpha^i$ being the initial scaling factor and $\alpha^f$ the final one. We may consider different ranges for MHSA and FFNs, since they follow different patterns. $b$ is the block index of the submodel and $B$ represents the total number of blocks in the submodel. The rank $r_l$ of layer $l$ is computed by $r_l = \alpha_l \cdot \min\{m, n\}$. The procedure is shown in Algorithm \ref{alg:linear_rank}.

\vspace{-0.2cm}
\begin{algorithm}
\caption{Linear Layer-Wise Rank Training from Scratch}
\label{alg:linear_rank}
\begin{algorithmic}
\scriptsize
\State \textbf{Inputs:} Full-rank model $\mathcal{M}$, set of factorized layers $\mathcal{S}$; scaling factors $\Gamma = [\alpha^{i_{MHSA}}, \alpha^{f_{MHSA}}, \alpha^{i_{FFN}}, \alpha^{f_{FFN}}$].

\For{$l \in \mathcal{S}$}
    \State $\alpha^i, \alpha^f, b, B$ = getLayerParameters($\mathcal{M}$, $l$, $\Gamma$)
    \State Compute $\alpha_l$ following Equation (\ref{eq:linear_rank}) with parameters $\alpha^i, \alpha^f, b, B$
    \State Compute rank $r_l = \alpha_l \cdot \min\{m, n\}$
    \State Initialize $U_l\in \mathbb{R}^{m \times r_{l}}$ and $V^T_l \in \mathbb{R}^{r_{l} \times n}$ with SVD following Eq. (\ref{eq:svd_init}) 
    \State Remove weight matrix $W_l$ 
\EndFor
\State Train the model \\
\Return Low-Rank Model $\rightarrow$ $\mathcal{M'}$
\end{algorithmic}
\end{algorithm}
\vspace{-0.3cm}

%% file: figures/architecture.tex
\begin{figure}[tb]
  \centering
  \captionsetup{font=scriptsize}
  \includegraphics[width=\columnwidth, trim=2.1cm 1.1cm 9.6cm 6.6cm, clip]{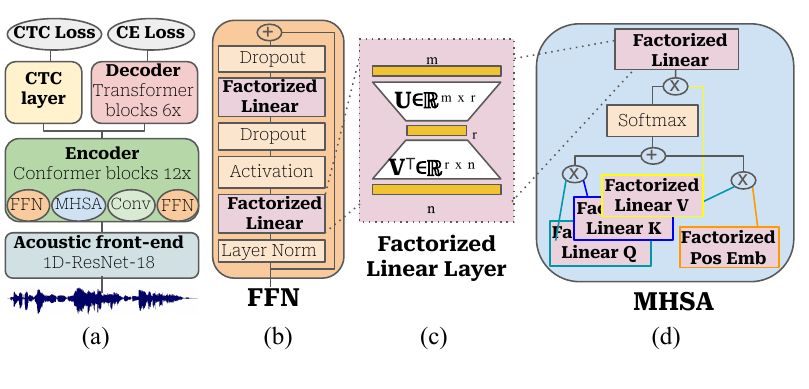}
  \vspace{-1cm}
  \captionsetup{font=scriptsize}
  \caption{(a) End-to-end ASR architecture. (b) Feed-Forward Network. (c) Factorized Linear Layer. (d) Multi-Headed Self-Attention.}%
  \label{fig:architecture}
  \vspace{-0.5cm}
\end{figure}

%% file: figures/linear_mapping.tex
\begin{figure}[tb]
  \centering
  \captionsetup{font=scriptsize}
  \includegraphics[width=\columnwidth, trim=0.5cm 3.5cm 5cm 1cm, clip]{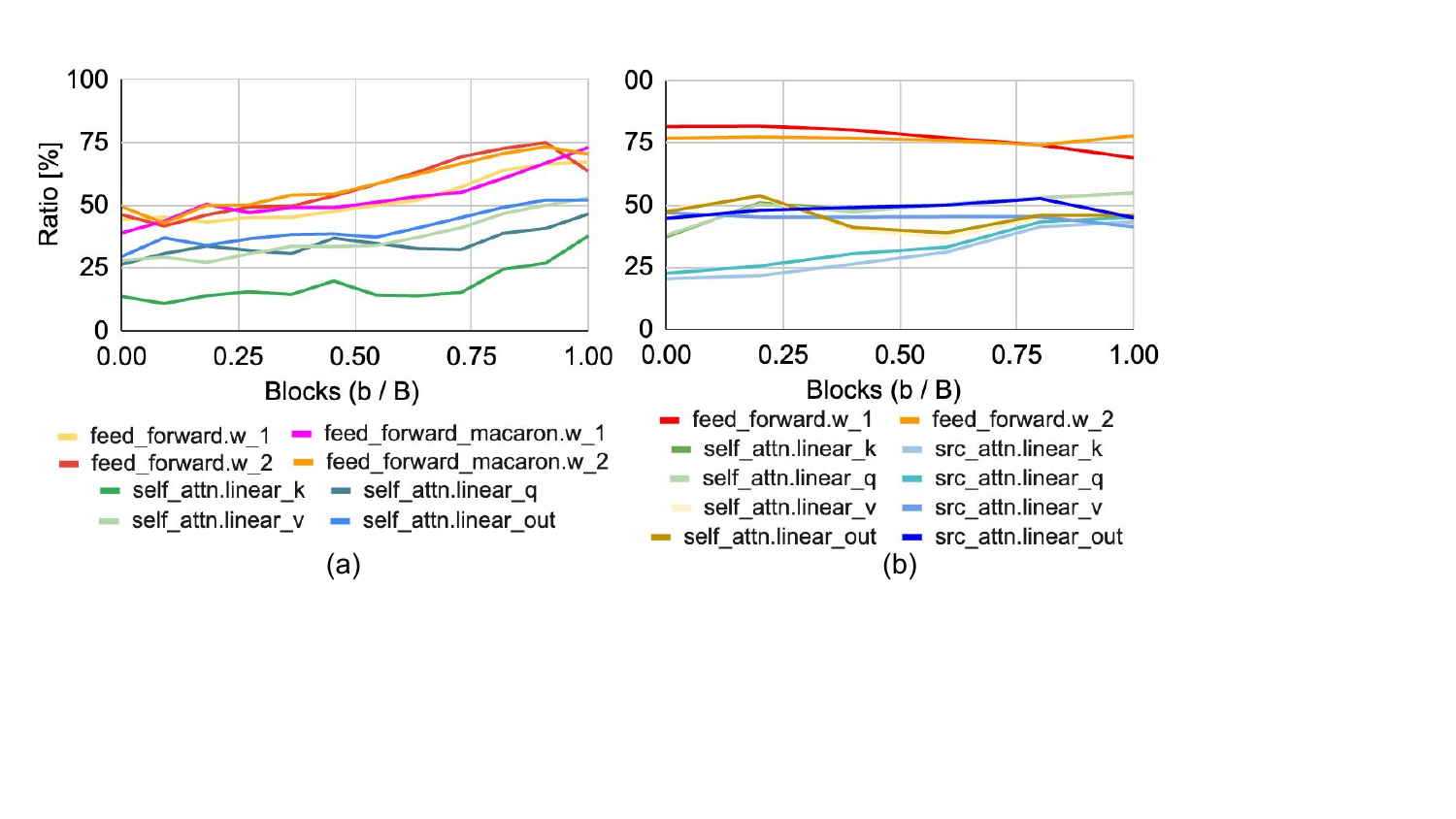}
  \vspace{-0.9cm}
  \captionsetup{font=scriptsize}
  \caption{(a) Conformer encoder blocks. (b) Transformer decoder blocks. Ratio: number of singular vectors required to approximate 95\% of each weight matrix compared to the total number of singular values. 
  The X-axis represents the model depth from top to end. $b$ is the current block and $B$ the number of blocks. A ratio close to 100\% indicates that all singular vectors are necessary, while a ratio close to 0\% means none are needed.}
    \label{fig:linear_mapping}
  \vspace{-0.6cm}
\end{figure}

%% file: sections/experimental_setup.tex
\section{Experimental setup}
\label{sec:experimental_setup}

\noindent\textbf{Dataset.}\quad
For ASR, we conduct experiments on two datasets: Librispeech~\cite{panayotov2015librispeech} and LRS3~\cite{afouras2018lrs3}. For Librispeech, we use ``train-clean-100'', ``train-clean-360'', and ``train-other-500'' subsets, totalling 960 hours of training data and evaluate our performance on the ``test-clean'' set with a total of 5.1 hours of audio. LRS3 consists of 439 hours of video clips, with~118,516~(408 hours),~31,982~(30 hours) and~1,321~(0.9 hours) clips in the pre-training, training-validation, and test sets, respectively. For AVSR, we use the same LRS3 splits for training and test. Additionally, following previous works \cite{ma2023auto, fernandez2023sparsevsr, fernandez2024msrs}, we use VoxCeleb2 \cite{chung2018voxceleb2} and AVSpeech \cite{ephrat2018looking} audio-visual datasets for training, resulting in 1,307 and 1,323 hours. We report results using Word Error Rate (WER). 

For ASR, we only apply adaptive time masking~\cite{ma2022visual} to the raw audio stream. Specifically, we select a number of masks that is proportional to the length of the utterance and a maximum masking duration of up to 0.4 seconds. For AVSR, in addition to adaptive time masking, we also apply horizontal flipping and random cropping.

\noindent\textbf{Pre-processing.}\quad For ASR, 
raw audio waveforms are used as input to the model and undergo z-normalization per utterance before being fed into the model \cite{DBLP:journals/corr/abs-2102-06657}. For AVSR, in addition, we crop a 96$\times$96 region centered around the mouth and then convert each frame into a greyscale image \cite{ma2022visual}.

\noindent\textbf{Model architecture.}\quad Instead of focusing on achieving state-of-the-art performance, our main objective is to explore low-rank training from scratch. To this end, we adapt the open-source architectures presented in~\cite{ma2023auto}. Our ASR model comprises a 1D ResNet front-end, followed by a Conformer encoder, a Transformer decoder and a CTC layer, resulting in 243M parameters. Following \cite{fernandez2024msrs}, our AVSR model comprises a ResNet frontend for audio and video modalities, a multi-layer perceptron for early fusion of multi-domain features, a single Conformer encoder, a Transformer decoder and a CTC layer, resulting in 268.5M parameters.

\noindent\textbf{Training details.}\quad
The models are trained for 75 epochs using the AdamW optimiser~\cite{loshchilov2017decoupled}. A constant learning rate with square root cooldown scheduler \cite{hagele2024scaling} is used for ASR, while a cosine scheduler is used for AVSR. The peak learning rate is 0.0006/0.001 for ASR/AVSR, 15 epochs of cooldown for ASR and a warm-up of 5 epochs for both. The ASR/AVSR models are trained with 32/64 A100 GPUs, respectively.

%% file: sections/results.tex
\vspace{-0.1cm}
\section{Results}
\label{sec:results}
\vspace{-0.1cm}

\input{tables/uniform_rank}

\noindent\textbf{Low-Rank Initialization Comparison.}\quad 
Figure \ref{fig:initializations} illustrates the performance of low-rank ASR models initialized using two different methods: the traditional Kaiming initialization \cite{he2015delving} (explored before for relatively small speech models \cite{hernandez2023sharing, winata2020lightweight}) and SVD initialization (as detailed in Section \ref{sec:svd_init}). In this analysis, uniform low-rank training was applied across all FNNs. The results clearly demonstrate that SVD initialization yields more stable training dynamics and minimizes performance degradation across varying scaling factors $\alpha$. These findings underline the efficacy of SVD initialization, prompting its adoption in all subsequent experiments.

\input{figures/initializations}

\noindent\textbf{Layer Factorization Experiments.}\quad In Table \ref{tab:uniform_rank}, we present a comparison of low-rank ASR models with different layer factorization strategies. We examine three scenarios: i) only linear layers in MHSA blocks are factorized; ii) only linear layers in FFNs are factorized; iii) both linear layers in MHSA blocks and FFNs are factorized.

Our experiments yield several notable insights. \underline{First}, we observe that uniformly reducing the rank of the FFNs results in a significant decline in model's performance. For example, reducing the feature dimension by 50\%, which leads to an approximate 20\% reduction in model size, results in an absolute increase of 0.41\% in WER on LRS3 and 0.26\% in WER on Librispeech. This indicates that FFNs are more resistant to low-rank factorization. \underline{Second}, we find that the linear layers in MHSA blocks can be effectively factorized from scratch with minimal performance degradation. Interestingly, the application of low-rank structures in these layers can even improve performance. For instance, applying a scaling factor of 0.12 to the linear layers in MHSA blocks improved the ASR model’s performance compared to the baseline (1.92\% WER vs. 1.99\% WER on LRS3), and similarly on Librispeech (2.61\% vs. 2.66\%). \underline{Third}, applying a uniform low-rank structure to both the MHSA and FFN linear layers led to unstable training, often resulting in exploding gradients and model degradation.

Note that our objective is to reduce model size while maintaining performance. While factorizing the MHSA layers can slightly enhance performance, it does not yield substantial memory reductions or speed-ups, as the majority of parameters are located in the FFN layers. Conversely, factorizing the FFN layers results in smaller models, but at the cost of performance loss. These findings suggest that applying a uniform low-rank pattern across all layers is suboptimal, and a more tailored strategy is required for efficient low-rank training. 

\input{tables/linear_rank}
\noindent \textbf{Layer-Wise Rank Comparison.}\quad Figure \ref{fig:linear_mapping} illustrates that the blocks of the Conformer and Transformer models exhibit different behaviors based on their location in the model. Specifically, early blocks are more amenable to low-rank approximation, while later blocks are less suitable for low-rank approximation. Additionally, MHSA layers tend to have lower rank than FFN layers (around a 25\% lower rank), which is consistent with the findings from the previous section. 
In Table \ref{tab:linear_rank}, we investigate a linear scaling factor that increases the rank as the depth of the model increases (as explained in Section \ref{sec:linear_rank}). Specifically, we examine multiple low-rank projections of the ASR model on both LRS3 and Librispeech. Our findings indicate that a linear mapping can significantly reduce the number of model parameters while maintaining the model's performance. Notably, reducing the model size by more than 50\% (115 M parameter model) results in a model that is 1.3$\times$ faster and uses around 10\% less memory, without compromising its performance. This was not observed when using a uniform rank, as a comparable model (113 M) requires larger batch size and sometimes results in unstable training. Additionally, it increases WER by 0.48\% on LRS3 and 0.17\% on Librispeech. 
Therefore, we can effectively low-rank linear layers by assigning an appropriate ratio based on their location within the network. Furthermore, when the ranks are accurately assigned based on both layer type and depth, there is no need for a warm-up period. 

\noindent \textbf{Low-Rank Training for AVSR.} Table \ref{tab:avsr_rank} showcases a comparison of various low-rank AVSR models evaluated on LRS3. 
Our linear rank mapping with 130 M parameters stands out for its ability to reduce parameters count by 50\%, leading to a notable 1.15$\times$ speed-up and 10\% memory savings. Notably, this is achieved while maintaining performance levels, with only a slight degradation observed in extremely noisy scenarios (0.8\% WER increment at a SNR of -7.5 dB).
\input{tables/avsr}

%% file: tables/uniform_rank.tex
\begin{table}[tb]
\setlength{\tabcolsep}{1.6pt}
\centering
\captionsetup{font=scriptsize}
\caption{WER [\%] ($\downarrow$) of ASR models trained from scratch with different uniform scaling factors on the test sets of LRS3 and Librispeech. Training speed-ups and memory costs on a maximum number of frames per batch of 2400, using uniform layer-wise rank. $^*$Unstable training, needs larger batch size (trained on 64 GPUs).}
\vspace{-0.1cm}
\resizebox{\columnwidth}{!}{
\begin{tabular}{ccccccc}
\toprule
\multirow{2}{*}{\textbf{Factorized Layers}} & \multirow{2}{*}{\textbf{$\alpha$}} & \multirow{2}{*}{\textbf{\# Params}}  & \multirow{2}{*}{\textbf{Speed-up}} & \multirow{2}{*}{\textbf{Memory [GB]}} & \multicolumn{2}{c}{\textbf{WER [\%]}}  \\ \cline{6-7}
 &  &      &  & & \textbf{LRS3}      & \textbf{Librispeech} \\ \toprule
\makecell{None (Full-Rank)}     &      N/A &                                            
           243 M            &          1      &  16.4 & 1.99      & 2.66 \\ \midrule
\multirow{3}{*}{MHSA \& FNNs}                                                    &
$^*$0.25                  &   113 M            & 1.36$\times$ & 14.6& 2.47 & 2.83\\ 
&$^*$0.23                  &   107 M                  &   1.36$\times$      & 14.5 &  2.27    &   2.87 \\
&0.12                  &   74.2 M                   & 1.45$\times$     &   13.9 &  2.48     &   3.03 \\ \midrule
\multirow{3}{*}{MHSA} & 0.25                  & 211 M                     &  1.03$\times$       & 16.0& 2.50      & 2.84\\
& 0.12                 & 194 M                    &  1.06$\times$      & 15.6 &  1.92      & 2.61 \\
& 0.06                  &    186 M                         &   1.06$\times$      &  15.5 & 1.92      &  2.73\\\midrule
\multirow{4}{*}{FFNs}                                                     &
0.50                   & 189 M                       &  1.15$\times$     & 15.8&  2.40      & 2.92\\ 
&0.25                  & 145 M                    &   1.33$\times$      & 15.0 & 2.41      & 2.97 \\
&0.06                  & 112 M                     &    1.41$\times$     & 14.4 & 2.48      &    3.07\\ \bottomrule 
\end{tabular}
}
\label{tab:uniform_rank}
\vspace{-0.4cm}
\end{table}

%% file: figures/initializations.tex
\begin{figure}[ht]
  \centering
  \includegraphics[width=0.75\columnwidth, trim=1cm 9.5cm 14cm 1cm, clip]{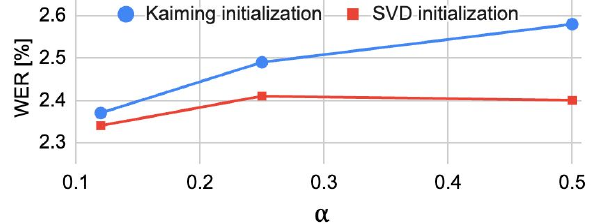}
  \captionsetup{font=scriptsize}
  \vspace{-0.1cm}
  \caption{WER [\%] ($\downarrow$) of low-rank ASR models trained from scratch on LRS3. The models are initialized with either SVD or Kaiming \cite{hernandez2023sharing, winata2020lightweight}. We compare their performance when different uniform scaling factors $\alpha$ are applied to FFNs.}
  \label{fig:initializations}
  \vspace{-0.2cm}
\end{figure}

%% file: tables/linear_rank.tex
\begin{table}[tb]
\setlength{\tabcolsep}{1.2pt}
\centering
\captionsetup{font=scriptsize}
\caption{WER [\%] ($\downarrow$) of ASR models trained from scratch with different linear scaling factors on the test sets of LRS3 and Librispeech. Training speed-ups and memory costs on a maximum number of frames per batch of 2400. $^*$Unstable training, needs a larger batch size to be stable, i.e., using 64 GPUs.} 
\resizebox{\columnwidth}{!}{
\begin{tabular}{lcccccc}
\toprule
\multirow{2}{*}{\textbf{\# Params}} & \textbf{MHSA} & \textbf{FNNs}  & \multirow{2}{*}{\textbf{Speed-up}} & \textbf{Memory} & \multicolumn{2}{c}{\textbf{WER [\%]}} \\   \cline{6-7}
& [\textbf{$\alpha^i$}, \textbf{$\alpha^f$}] & [\textbf{$\alpha^i$}, \textbf{$\alpha^f$}] & &  \textbf{[GB]} & \textbf{LRS3} & \textbf{Librispeech} \\\midrule
243 M            & \multicolumn{2}{c}{Full-Rank}         &        1      &  16.4 & 1.99      & 2.66  \\ \midrule
113 M &  \multicolumn{2}{c}{$^*$$\alpha=0.25$}   & 1.36$\times$ & 14.6& 2.47 & 2.83\\ 
107 M & \multicolumn{2}{c}{$^*$$\alpha=0.23$}   & 1.36$\times$ & 14.5& 2.27    &   2.87\\ 
74.2 M & \multicolumn{2}{c}{$\alpha=0.12$}            & 1.45$\times$        & 13.9 &2.48     &   3.03\\  \midrule
197 M               &        [0.1,      0.2]            &       Full-Rank                            &    1.04$\times$              & 15.7&    1.98    &        2.52 \\
161 M               &        [0.2,      0.4]            &       [0.4,      0.6]                     &    1.16$\times$              & 15.5 &      2.00    &           3.35\\
115 M                    &        [0.1,       0.2]           &          [0.2,      0.5]                            &   1.31$\times$               & 14.9&   2.11          &     2.74  \\  
107 M                   &        [0.1,      0.2]           &          [0.2,      0.4]                          &   1.32$\times$               & 14.4 &         2.07      &        2.80  \\
\bottomrule 
\end{tabular}
}
\label{tab:linear_rank}
\vspace{-0.5cm}
\end{table}

%% file: tables/avsr.tex
\vspace{-0.1cm}
\begin{table}[thb]
\setlength{\tabcolsep}{1.2pt}
\captionsetup{font=scriptsize}
\caption{WER [\%] ($\downarrow$) of AVSR models as a function of the noise levels on the LRS3 test set. Models are trained with different linear scaling factors on a combination of LRS3, VoxCeleb2, and AVSpeech datasets in the presence of babble noise from NOISEX \cite{varga1993assessment}. }
\centering
\resizebox{\columnwidth}{!}{
\begin{tabular}{ccccccccccc}
\toprule
\textbf{Layer-wise}& \multirow{2}{*}{\textbf{$\#$Params}} & \textbf{MHSA} & \textbf{FFNs} & \multicolumn{6}{c}{\textbf{SNR [dB]}} \\  \cline{5-10}
\textbf{rank} &  & [\textbf{$\alpha^i$}, \textbf{$\alpha^f$}] & [\textbf{$\alpha^i$}, \textbf{$\alpha^f$}] & \textbf{-7.5} & \textbf{-2.5} & \textbf{2.5} & \textbf{7.5} & \textbf{12.5} & \textbf{Clean} \\ \midrule
Auto-AVSR \cite{ma2023auto}                       & 443 M     &\multicolumn{2}{c}{Full-Rank}                               & 5.6           & 2.2           & 1.5          & 1.0          & 1.0           & 0.9          \\ 
MSRS \cite{fernandez2024msrs}                           & 268 M  &\multicolumn{2}{c}{Full-Rank} & 2.8          &  1.5          & 1.1         & 1.0         & 0.9 & 0.8          \\ \midrule
Linear rank & 156 M &        [0.1,      0.3]            &       [0.3,      0.5]           & 4.5 & 2.3 & 1.6 & 1.3 & 1.3 & 1.3 \\ 
Linear rank & 140 M &        [0.1,      0.2]            &       [0.2,      0.5]           & 3.9 & 2.1 & 1.4 & 1.3 & 1.2  & 1.2 \\ 
Linear rank &  132 M &        [0.1,      0.2]            &       [0.2,      0.4]           & 3.6 & 1.8 & 1.3 & 1.0  & 0.9  & 0.9 \\ \bottomrule
\end{tabular}
}
\label{tab:avsr_rank}
\end{table}

%% file: sections/conclusions.tex
\vspace{-0.4cm}
\section{Conclusions}
\label{sec:conclusions}
\vspace{-0.1cm}
This study explores the emergence of low-rank structures during the training of speech recognition models and uncovers two key findings: (i) as training progresses, the entire model gradually adopts a low-rank structure, with certain layers consistently exhibiting lower ranks than others; (ii) early blocks tend to have lower ranks compared to later blocks, following a near-linear pattern.
We found that SVD initialization and linear layer-wise rank enhance low-rank training. By leveraging these techniques, we propose LR-SMS, an effective linear rank mapping that enables the training of large-scale speech models with low-rank weights from scratch. LR-SMS achieves substantial reductions in parameter count, memory usage and training time while matching the performance of full-rank training, making it a promising solution for efficient and scalable speech recognition.